\documentclass[%
 reprint,
 amsmath,amssymb,
 prb,
]{revtex4-1}

\usepackage{graphicx}
\usepackage{dcolumn}
\usepackage{bm}

\begin{document}

\preprint{APS/123-QED}

\title{Metallic nanolines ruled by grain boundaries in graphene: an {\it ab 
initio}
study}

\author{F. D. C. de Lima, and R. H. Miwa}
\affiliation{
   Instituto de F\'isica, Universidade Federal de Uberl\^andia,
        C.P. 593, 38400-902, Uberl\^andia, Brazil}

\date{\today}

\begin{abstract}
 
We have performed an {\it ab initio}  investigation of the energetic stability, 
and the electronic properties of transition metals (TMs = Mn, Fe, Co, and Ru) 
adsorbed on  graphene upon the presence of grain boundaries (GBs). Our  results 
reveal an  energetic preference for the TMs lying  along the GB sites (TM/GB). 
Such an energetic preference has been strengthened by increasing the 
concentration of the TM adatoms; giving rise to  TM nanolines on graphene ruled 
by  GBs. Further diffusion  barrier calculations for Fe adatoms support the 
formation of those TM nanolines. We find that the energy barriers parallel to 
the GBs  are  sligthly lower in comparision with those obtained for the defect 
free graphene; whereas, perpendicularly to the GBs the Fe adatoms face higher 
energy barriers. Fe and Co (Mn) nanolines are  ferromagnetic (ferrimagnetic), in 
contrast the magnetic state of Ru nanolines is sensitive to the Ru/GB adsorption 
geometry. The electronic properties of those TM nanolines were characterized 
through extensive  electronic band structure calculations. The formation of 
metallic nanolines is mediated by a strong hybridization between the TM and the 
graphene ($\pi$)  orbitals along the GB sites. Due to the net magnetization of 
the TM nanolines,  our band structure results indicate an  anisotropic 
(spin-polarized) electronic current for some TM/GB systems.


\end{abstract}

\pacs{Valid PACS appear here}
\maketitle


\section{Introduction}

Topological defects in two dimensional systems has been the subject of several 
studies addressing not only their intrinsic properties, but also the use of 
those structural defects to perform nanoengineering on 2D platforms. Currently, 
graphene has been the most studied  2D platform; where we may have localized 
topological defects  like Stone-Wales\,\cite{choiPRL2000,meyerNanoLett2008} and 
self-interstitials\,\cite{luskPRL2008,lehtinenNanoLett2015}, or extended defects 
like grain boundaries\,\cite{simonisSUSC2002,yazyevPRB2010,huangNature2011}.

Grain boundaries (GBs) may change the electronic transport properties in 
graphene\,\cite{yazyevNatMat2010,mendezNanoscale2011}. Very recently,  extended 
defects have been synthesized in a controlled way, giving rise to (i) metallic 
channels embedded in graphene\,\cite{lahiriNatNanotech2010}. Meanwhile, (ii) 
ferromagnetic properties have been predicted along the  GB sites in graphene,  
upon external tensile strain\,\cite{kouACSNano2011} or  n-type 
doping\,\cite{alexandreNanoLett2012}. Further theoretical studies indicate that 
(iii) through a suitable incorporation of nitrogen atoms along the GB sites, we 
may have an electronic confinement effects, giving rise to semiconductor 
channels in graphene\,\cite{britoNanotech2014}. Those properties [(i)--(iii)] 
depend on the atomic geometry along the GBs.

 The presence of reconstructed defects enhances the chemical reactivity of the 
GB sites\,\cite{cretuPRL2010}. Indeed, GBs in graphene have been identified  by 
deposition of foreign elements, like silver\,\cite{kebaliEPhysJD2009} or 
gold\,\cite{yuACSNano2014}. Such an enhanced reactivity can be used to mediate 
(self) assembly processes on graphene, giving rise to nanoline (NL) strutures 
ruled by GBs. Those NLs may provide a set of new/useful electronic and 
chemical properties. In a recent experimental work\,\cite{kimNatComm2014}, the 
authors verified that the performance of hydrogen gas sensors was improved, upon 
the formation of linear structure of Pt adatoms along the GB sites. Further 
theoretical studies indicate that NLs of   Fe and Mn adatoms lying  along 
the GBs, composed by a pair of pentagons and an 
octagon\,\cite{lahiriNatNanotech2010}, give rise to half-metallic nature for the 
electronic transport along the Fe or Mn decorated GB 
sites\,\cite{zhuJPhysD2014,obodoPRB2015}. Indeed, somewhat similar 
spin-polarized current, dictated by the presence of foreign atoms, has been 
predicted along the edge sites of graphene 
nanoribbons\,\cite{martinsPRL2007,rigoPRB2009}.

In this work we have performed an {\it ab initio} study, based on the density 
functional theory (DFT), of the energetic stability and the electronic 
properties of transition metals (TMs = Mn, Fe, Co, and Ru) adsorbed on graphene 
upon the  presence of grain boundaries.  Here we have considered  a 
number of plausible TM--GB configurations for  two different GB geometries. We 
find an energetic preference for the TMs lying along the GB sites (TM/GB), 
giving rise to TM nanolines (TM-NLs) ruled by GBs. Due to the metal--metal 
(chemical) interactions, the energetic stability of the TM/GB structures 
has been strengthened by increasing the concentration of TM adatoms along 
the GB sites. Those results provide further support to the  recent experimental 
findings of  linear structures of TMs  on graphene, patterned by 
GBs\,\cite{kebaliEPhysJD2009,yuACSNano2014,kimNatComm2014}. In addition, due to 
the  net magnetic moment of the TM adatoms,  we find that 
the Fe- and Co-NLs are ferromagnetic, while Mn-NLs are 
ferrimagnetic. In contrast, the magnetic state (ferromagnetic/nonmagnetic) of  
Ru-NLs is sensitive to the (local) adsorption geometry.   Our electronic band 
structure calculations reveal that  the (most of) TM/GB systems are 
metallic, and indicate a spin-anisotropy for the electronic current along the 
TM-NLs.

\section{Method}

The calculations were performed based on the DFT approach, as implemented in the
VASP code\,\cite{vasp1}. The exchange correlation term was described by using
the spin-polarized GGA approach, in the form proposed by Perdew, Burke and
Ernzerhof\,\cite{PBE}. The Kohn-Sham orbitals are expanded in a plane
wave basis set with an energy cutoff of 400\,eV. The 2D Brillouin Zone (BZ) is
sampled according to the Monkhorst-Pack method\,\cite{mp} (MP), using
a 8$\times$8$\times$1 mesh for GB(5-8) and 6$\times$6$\times$1 mesh for GB(5-7).
We made additional convergence tests with respect
to the energy cutoff (up 450\,eV) and BZ sampling (MP mesh of up to 
20$\times$20$\times$1). The
electron-ion interactions are taken into account using the Projector Augmented
Wave (PAW) method \cite{kressePRB1999}. All geometries have been relaxed
until atomic forces were lower than 0.025~eV/\AA. The molecule/graphene system
is simulated using the slab method, by considering a vacuum region in the
direction perpendicular to graphene sheet of at  least 8\,\AA.

\section{Results}

\subsection{TM/Graphene and Pristine GBs}

\begin{figure}[h]
 \includegraphics[width= 8.5cm]{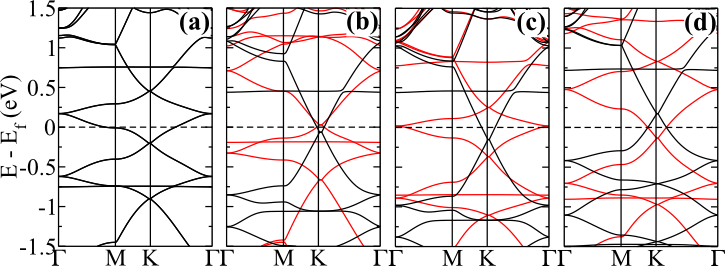}
 \caption{\label{graph} Electronic band structure of TM adsorbed 
graphene layer, TM/graphene,  for TM = Mn (a), Fe (b), Co (c), and Ru (d). The 
Fermi level was set to zero.}
 \end{figure}

Initially we examine the energetic stability and the electronic properties of 
the TMs (TMs = Mn, Fe, Co, and Ru) adsorbed on the pristine graphene layer. 
The adsorption energy $(E^a)$ can be written as, 
$$
E^a = E[{\rm graphene}] + E[{\rm TM}] - E[{\rm TM/graphene}],
$$
where $E[{\rm graphene}]$ and $E[{\rm TM}]$ are the total energies of the 
separated components, graphene sheet and TM atom, and $E[{\rm TM/graphene}]$ is 
the total energy of the final system, namely graphene sheet adsorbed by TM, 
TM/graphene. In Table~I we present our results of $E^a$, net magnetic moment 
(m), and the nearest neighbor (NN) TM--C  equilibrium  distances for the TMs 
adsorbed on the (energetically more stable) hollow sites of graphene. Here, we 
find a good agreement with the previous theoretical studies, {\it viz.}: $E^a$ = 
0.17\,eV/Mn-atom\,\cite{sevincliPRB2008}, $E^a$ = 1.02, 0.85, and 
0.65\,eV/Fe-atom\,\cite{sevincliPRB2008,chanPRB2008,wangJPhysCondMatt2009}, 
$E^a$ = 1.27\,eV/Co-atom\,\cite{sevincliPRB2008}, and $E^a$ = 
2.64\,eV/Ru-atom\,\cite{acostaPRB2014}. The electronic band structures of 
TM/graphene are presented in Fig.~\ref{graph}. The linear energy dispersion at 
the  K and K' points have been preserved, however,  for Fe/, Co/, and 
Ru/graphene systems the  Dirac cones exhibit a spin-split due to the 
exchange-field induced by the adatoms. 
%

 \begin{figure}[h]
 \includegraphics[width= 8.5cm]{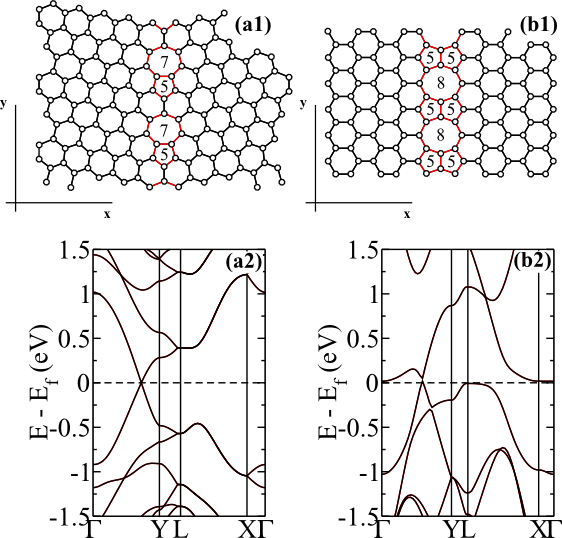}
 \caption{\label{pristine} Structural models and the electronic band structure 
of GBs, GB(5-7) (a) and GB(5-8) (b). The 
Fermi level was set to zero.}
 \end{figure}


\begin{table}[h]
\caption{Adsoption energy ($E^a$ in eV/atom), TM--C equilibrium bond length (in 
\AA), and the net magnetic moment (m in $\mu_B$) of TMs adsorbed on the hollow 
site of pristine graphene.}
\begin{ruledtabular}
\begin{tabular}{cccc}
TM & $E^a$ & TM--C & m  \\
\hline
Mn & 0.19  &   2.08  & 0.00  \\
Fe & 0.73  &   2.11  & 2.01  \\
Co & 1.14  &   2.11  & 1.16 \\
Ru & 2.41  &   2.25  & 1.46  \\
\end{tabular}
\end{ruledtabular}
\end{table}

In Fig.~\ref{pristine} we present the structural models of GBs in graphene, and 
their electronic band structures. In GB(5-7) we have a defect line composed by 
carbon pentagons and heptagons embedded in the graphene sheet 
[Fig.\,\ref{pristine}(a1)], whereas GB(5-8) is composed by two pentagons and an 
octagon along the graphene zigzag direction [Fig.\,\ref{pristine}(b1)]. The 
electronic  band structures  are characterized by a linear energy dispersion,
near the Dirac Point (DP), for wave vectors parallel to the defect lines, 
Figs.\,\ref{pristine}(a2) and \ref{pristine}(b2). Scanning tunelling microscopy 
(STM) experiments show the  formation of bright lines along the GBs in 
graphene\,\cite{simonisSUSC2002,lahiriNatNanotech2010,huangNature2011} .  Such 
STM picture has been supported by recent first-principles 
simulations\,\cite{yazyevPRB2010}, and indicates that the electronic states near 
the Fermi level ($\rm E_F$) are mostly localized along the GBs. The increase of 
the electronic density of states near $\rm E_F$ suggests a more reactive 
character of the  GB sites in comparison with the pristine region of the 
graphene sheet\,\cite{britoAPL2011}. Indeed, our $E^a$ results reveal that there 
is an energetic preference for TMs adsorbed along the GB sites of GB(5-7) and 
GB(5-8). 

 \begin{figure*}
 \includegraphics[width= 16cm]{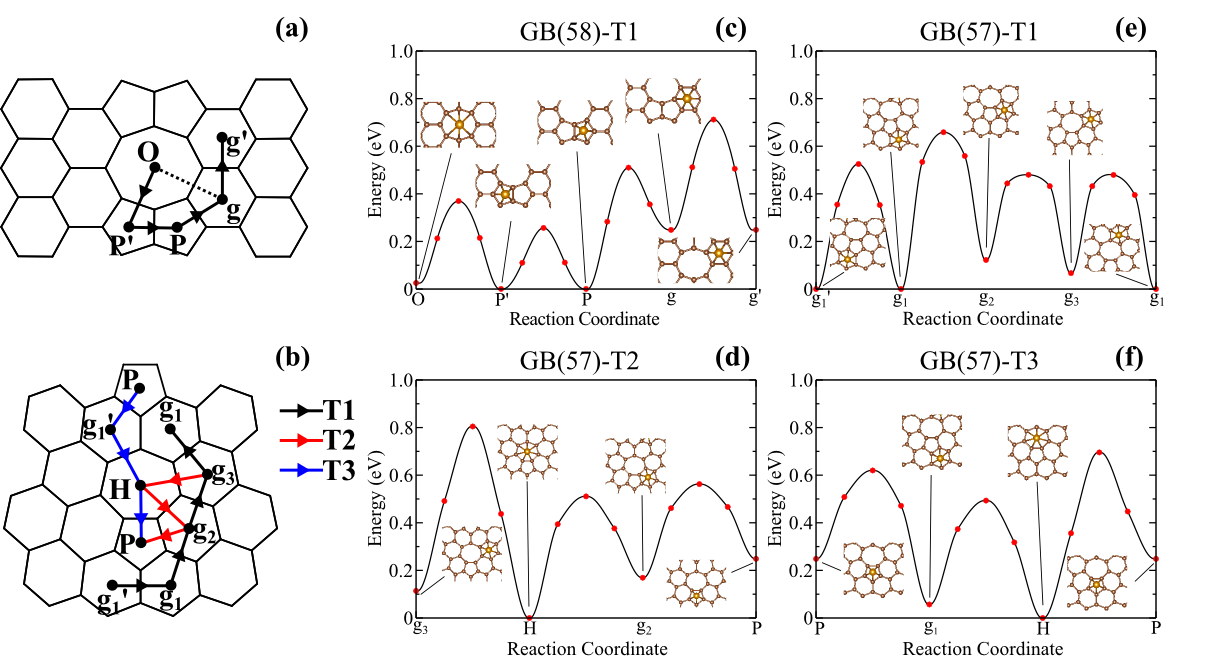}
 \caption{\label{neb} 
Diffusion paths for Fe adatoms in GB(5-8) (a) and 
GB(5-7); and the calculated energy barriers of Fe/GB(5-8) along the diffusion 
path T1 (c), and the energy barriers of Fe/GB(5-7) along the diffusion paths T1 
(d), T2 (e), and T3 (f).}
 \end{figure*}

\begin{table}[h]
\caption{Adsoption energy ($E^a$ in eV/atom), and the net magnetic moment (m in
$\mu_B$) of TM/GB(5-8) and TM/GB(5-7) systems, for the TM adsorbed on the 
pentagonal (P), heptagonal (H), and octagonal (O) hollow sites.}
\begin{ruledtabular}
\begin{tabular}{ccccc} 
\multicolumn{1}{c}{TM/GB(5-8)} & 
\multicolumn{2}{c}{$E^a$} &
\multicolumn{2}{c}{m}\\
\cline{2-3}\cline{4-5}
          &  P     &   O   &   P       & O \\
\hline
Mn        &  1.31  & 1.16  &   4.30    & 4.27 \\
Fe        &  1.62  & 1.61  &   2.76    & 3.12 \\
Co        &  2.00  & 1.64  &   0.93    & 1.89 \\
Ru        &  3.63  & 2.81  &   0.00    & 1.70 \\
\hline   
\multicolumn{1}{c}{TM/GB(5-7)} & 
\multicolumn{2}{c}{$E^a$} &
\multicolumn{2}{c}{m}\\
\cline{2-3}\cline{4-5}
          &  P &  H & P  & H  \\
\hline
Mn         & 0.87  & 0.79   & 4.48 & 4.21 \\
Fe         & 1.06  & 1.36   & 2.54 & 2.13  \\
Co         & 1.52  & 1.53   & 1.03 & 1.05  \\
Ru         & 3.05  & 2.78   & 0.00 & 1.71 \\

\end{tabular}
\end{ruledtabular}
\end{table}

\subsection{One TM per GB unit cell}
 
We have considered a number of plausible configurations of TMs adsorbed along 
the GB sites (TM/GB); where we confirm the energetic preference of TM/GB, when 
compared  with the TMs adsorbed on the pristine graphene. For instance, we find 
that the Fe adatoms  lying on the pentagonal (P) rings of GB(5-8), 
Fe[P]/GB(5-8), are more strongly attached to the graphene sheet by around 
0.9\,eV/atom, $E^a = 0.73 \rightarrow  1.62$\,eV/atom. The same increase on the 
adsorption energy was verified for Co[P]/GB(5-8). For Mn[P]/ and Ru[P]/GB(5-8)  
the adsorption energy increases by about 1.2\,eV/atom. TMs adsorbed on the 
octagonal (TM[O]) ring is the second most stable configuration for Mn, Co and 
Ru, while Fe[P]/ and Fe[O]/GB(5-8) present almost the same adsorption energies. 
Our results of adsorption energies and the net magnetic moments are summarized 
in Table~II. The energetic preference for  Fe[P]/, and Co[P]/GB(5-8) are in 
agreement with the previous theoretical 
studies\,\cite{zhuJPhysD2014,yuJChemMatC2014,obodoPRB2015}, however, the same 
does not occur for the Mn adatom. We find an energetic preference  for the 
pentagonal sites, Mn[P]/GB(5-8)\,\cite{yuJChemMatC2014}, instead of 
Mn[O]/GB(5-8)\,\cite{zhuJPhysD2014,obodoPRB2015}. In order to verify the 
accuracy of our results, we have performed additional calculations  including up 
Mn-3s as valence orbitals for the Mn pseudopotential, (i) within the PAW 
approach~\cite{paw,kressePRB1999}, and (ii) using ultrasoft Vanderbilt 
pseudopotentials~\cite{vanderbilt}. In (i) we have used the VASP code, as 
described in Section II (increasing the energy cutoff to 450\,eV), and in (ii) 
we have used the Quantum-ESPRESSO code~\cite{espresso,MnGB}. In both cases [(i) 
and (ii)], the energetic preference for the P sites was confirmed, being the  
Mn[P]/GB(5-8) more stable than Mn[O]/GB(5-8) by 0.16 and 0.15\,eV, respectively.
 
The energetic preference for the GB sites has  been also verified in 
TM/GB(5-7); however, compared with the TM/GB(5-8) systems,  the adsorption 
energies are lower by 0.3--0.5\,eV/atom for the most stable configurations.
The adsorption energy results are also summarized in Table~II; where we find 
that  (i) the energetic preference for the pentagonal rings for Mn and Ru  
adatoms has been maintained, while (ii) Fe adatoms are more stable by 0.3\,eV on 
the heptagonal (H) ring, and  (iii) for Co adatoms, the P and H rings 
present almost the same adsorption energies. 

In particular, for  the Fe/GB systems, we examined the  Fe diffusion on the 
GB(5-8) and GB(5-7) defect lines. The energy barriers were estimated by  using 
the CI-NEB approach\,\cite{NEB}. For the Fe/GB(5-8) system, we have considered 
the diffusion path T1, depicted in Fig.\,\ref{neb}(a). We find that (i) the 
energy barrier ($\rm E^{barr} = 0.38$\,eV) for Fe diffusion along the GB sites 
passing through the C--C bridge site, Fe[O]$\rightarrow$Fe[P']$\rightarrow$Fe[O] 
or Fe[O]$\rightarrow$Fe[P']$\rightarrow$Fe[P]$\rightarrow$Fe[O] in 
Fig.\,\ref{neb}(c), is lower than  the one passing through hexagonal sites, 
Fe[g'] ($\rm E^{barr} =  0.45$\,eV); and (ii) the Fe adatom will face energy 
barrier of about 0.5\,eV to move out from the GB(5-8) defect line,  
Fe[P]$\rightarrow$Fe[g].  In Fe/GB(5-7), (iii) the Fe energy  barriers along the 
 hexagonal sites neighboring the GB, diffusion path T1 in Figs.\,\ref{neb}(b) 
and \ref{neb}(e), are slightly larger ($\rm E^{barr}$ between 0.5--0.6\,eV) 
compared with the one in GB(5-8); and (iv)  the energy  barriers from the 
hexagonal site g$_2$ to the H and P sites along T2 [Fig.\,\ref{neb}(d)], and 
from the hexagonal site g'$_1$ to the H site along T3 [Fig.\,\ref{neb}(f)] are 
slightly lower when compared with the energy barrier along T1, 
Fig.\,\ref{neb}(e). In contrast, (v) for the diffusion path 
Fe[H]$\rightarrow$Fe[P] (T3) we find  $\rm E^{barr}$ of 0.70\,eV, and thus 
suggesting  that  the Fe diffusion along the  GB(5-7) will take place passing 
through the neighboring hexagonal (g sites), for instance 
Fe[H]$\rightarrow$Fe[g$_2$]$\rightarrow$Fe[P] along T2. Those results suggest 
that the there is an energetic preference for the Fe diffusion along the GB 
sites of GB(5-8) [(i) and (ii)],  and GB(5-7) [(iv) and (v)]. Similar picture is 
expected for the other TM/GB systems.

\begin{figure*}
 \includegraphics[width= 18cm]{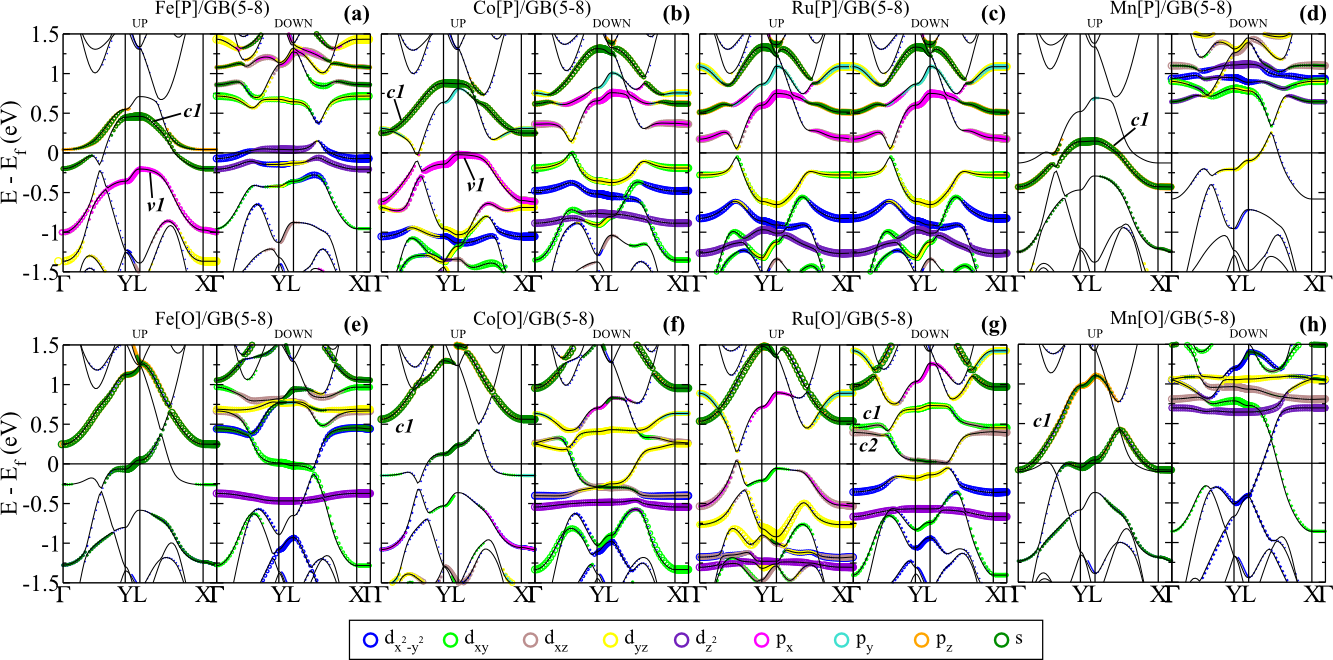}
 \caption{\label{banda-58} Electronic band structure of TM/GB(5-8) for 
TMs  adsorbed on the pentagonal  (P) site, TM[P]/GB(5-8),   TM[P] = Fe 
(a), Co (b), Ru (c), and Mn(d); and for  TMs adsorbed on the octagonal (O) 
site, TM[O]/GB(5-8), TM[O] = Fe (e), Co(f), Ru(g), and Mn (h). The Fermi level 
was set to zero.}
 \end{figure*}

\begin{figure*}
 \includegraphics[width= 18cm]{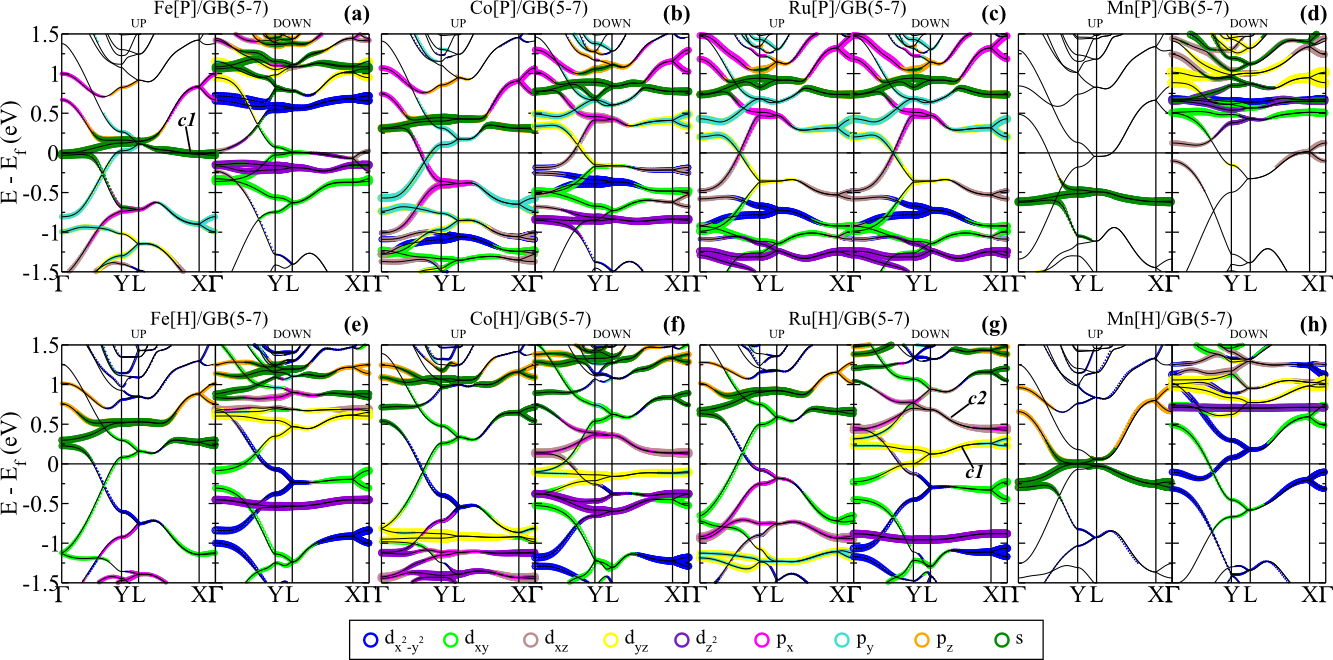}
 \caption{\label{banda-57} Electronic band structure of TM/GB(5-7) for 
TMs  adsorbed on the pentagonal  (P) site, TM[P]/GB(5-7),   TM[P] = Fe 
(a), Co (b), Ru (c), and Mn(d); and for  TMs adsorbed on the heptagonal (H) 
site, TM[H]/GB(5-H), TM[H] = Fe (e), Co(f), Ru(g), and Mn (h). The Fermi level 
was set to zero.}
 \end{figure*}

In GB(5-8) and GB(5-7), the electronic states near the Fermi level are mostly 
localized along the GB sites 
\cite{simonisSUSC2002,lahiriNatNanotech2010,yazyevPRB2010,huangNature2011}. Upon 
the TM adsorption, as shown in Figs.\,\ref{banda-58} and \ref{banda-57}, the 
electronic states of graphene hybridize with the ones of the TMs. For instance, 
in Fe[P]/GB(5-8) the spin-up channel of Fe-4s hybridizes with the host $\pi$ 
orbitals of graphene, giving rise to metallic states within $\rm E_F\pm0.5$\,eV, 
indicated as $c1$ in Fig.\,\ref{banda-58}(a). For the same spin-up channel, we 
find an occupied band ($v1$) just below $\rm E_F$,  with an energy dispersion of 
0.5\,eV, mostly composed by Fe-3p$_{\rm x}$ and graphene-$\pi$ orbitals. In 
contrast,  for the spin-down channel,  the Fe-3d$_{\rm x^2-y^2}$ and -3d$_{\rm 
z^2}$ orbitals give rise to dispersionless (flat) bands lying at the Fermi 
level. In general,   the electronic band structure of GB(5-8), near $\rm E_F$,  
has  been somewhat preserved for the spin-up channel of Fe[P]/GB(5-8), but not 
for the spin-down channel. Half-metal behavior has been proposed for Fe/ and 
Mn/GB(5-8)\,\cite{zhuJPhysD2014,obodoPRB2015,yuJChemMatC2014}. Here we show that 
the half-metal behavior in Fe[P]/GB(5-8) is ruled by a hybridization between the 
Fe-4s and the host $\pi$ orbitals [$c1$ in Fig.\,\ref{banda-58}(a)]. Fe[P] and 
Co[P]/GB(5-8) systems  present similar electronic band pictures near the Fermi 
level, however, the latter one does not present  half-metallic  band structure. 
In Co[P]/GB(5-8) the energy dispersion of $c1$  [Fig.\,\ref{banda-58}(b)] is 
practically the same as compared with the  Fe[P]/GB(5-8) system, however, it  is 
empty lying at $\rm \sim E_F + 0.5$\,eV. Meanwhile, the spin-up and -down energy 
bands are degenerated for Ru[P]/GB(5-8), Fig.\,\ref{banda-58}(c), where 
Ru-4d$_{\rm xy}$, -4d$_{\rm yz}$, and  hybridizes with the host graphene-$\pi$ 
orbitals at the Dirac cone. We find that there is no such a half-metal character 
for  Fe adatoms adsorbed on  the P sites of GB(5-7), Fig.\,\ref{banda-57}(a). 
The charge density overlap between the neighboring Fe adatoms is reduced in 
Fe[P]/GB(5-7).  Compared with  Fe[P]/GB(5-8), in Fe[P]/GB(5-7) there is an 
increase of the Fe[P]--Fe[P] distance between the pentagonal sites, 6.63\,\AA. 
In GB(5-8) the Fe[P]--Fe[P] distance is equal to 4.92\,\AA. The spin-down 
channel presents a set of  flat bands, near the Fermi level, mostly composed by 
Fe-3d$_{\rm  x^2-y^2}$ and -3d$_{\rm z^2}$ orbitals. As depicted in 
Figs.\,\ref{banda-57}(b) and \ref{banda-57}(c), (i)  the Dirac cone has been 
preserved, near the Fermi level,  for both spin channels of Co[P]/ and 
Ru[P]/GB(5-7), and (ii) the spin-up and -down bands are degenerated in 
Ru[P]/GB(5-7), as we obtained for Ru[P]/GB(5-8). Those results show that, 
although for both GBs the TMs are adsorbed on the P sites, we  find quite 
different electronic pictures; indicating that the the topology of the GB (host) 
sites plays a important role on the electronic properties of the TM/GB systems.
 
The half-metallic character of Fe[P]/GB(5-8) has been suppressed for Fe adatoms 
on the O sites, Fe[O]/GB(5-8) [Fig.\,\ref{banda-58}(e)];  we have metallic bands 
(composed by Fe-4s and -4d$_{\rm xy}$ orbitals) for both spin-channels. In 
Fe[O]/ and Co[O]/GB(5-8) [Fig.\,\ref{banda-58}(f)], the Dirac cone structure has 
 been preserved   for the spin-up channels, lying at $\sim$ $\rm E_F - 
0.3$\,eV. Meanwhile, the spin-down channels  are characterized  by, (i) 
electronic contribution of TM-3d states, hybridized with the host graphene-$\pi$ 
orbitals, to the formation of metallic bands for wave vectors parallel to the 
GBs ($\Gamma$Y and LX directions),  and (ii) localized TM-d$_{\rm z^2}$ orbitals 
at about $\rm E_F - 0.5$\,eV. We find that Ru[O] adatoms become spin-polarized. 
Ru[O]/GB(5-8) presents a net magnetic moment of 1.70\,$\mu_{\rm B}$ (mainly) due 
to the unoccupied spin-down 4d$_{\rm xy}$ and 4d$_{\rm xz}$ orbitals at $\rm E_F 
+ 1$\,eV, Fig.\,\ref{banda-58}(g). 

The electronic structures of Fe[H], Co[H]/ and Ru[H]/GB(5-7), 
depicted in Figs.\,\ref{banda-57}(e)--\ref{banda-57}(g), reveal that (i) the  
TM-d$_{\rm xy}$ and -d$_{\rm x^2 - y^2}$ states hybridizes with the 
graphene-$\pi$ orbitals, preserving the Dirac cone structure for the spin-up 
channels, within an energy window of $\rm E_F \pm 0.5$\,eV; whereas (ii) the 
spin-down channels are characterized by the presence of dispersionless TM-d  
orbitals near the Fermi level. Such an electronic structure gives rise to a 
spin-anisotropy in the electronic current along the GB sites of TM[H]/GB(5-7), 
for TM = Fe, Co and Ru. Ruled by (i) and (ii), the  spin-up current 
will be larger than that of spin-down; since the presence of localized states 
near the Fermi level, for the spin-down channel, may act as electron/hole 
scattering centers.

The electronic band structure along the $\Gamma$Y direction of  Mn[P]/ and 
Mn[O]/GB(5-8) [Figs.\,\ref{banda-58}(d) and \ref{banda-58}(h), respectively] 
suggest a half-metal character, in agreement with the previous theoretical 
studies\,\cite{zhuJPhysD2014,obodoPRB2015}. However,  along the LX direction of 
the spin-down channels,  we find metallic states composed by Mn-3d states  
hybridized with the graphene-$\pi$ orbitals, giving rise to a Dirac-type cone at 
about $\rm E_{F} + 0.3$\,eV. The formation such metallic bands, and the presence 
of a Dirac-type cone structure along the LX direction,  are in agreement 
with our previous studies of nitrogen (n-type) doped 
GB(5-8)\,\cite{britoNanotech2014}. Indeed, by comparing the electronic band 
structure of pristine GB(5-8) [Fig.\,\ref{pristine}(d)], and Mn/GB(5-8),  we 
verify a downshift of the DP, indicating a n-type doping of GB(5-8) upon the 
adsorption of both Mn[P] and Mn[O].  Thus, the presence of those metallic 
states rules  out the half metallic properties of Mn/GB(5-8). We believe that 
further investigations are necessary to clarify this  point.

 \begin{figure}
 \includegraphics[width= 8.5cm]{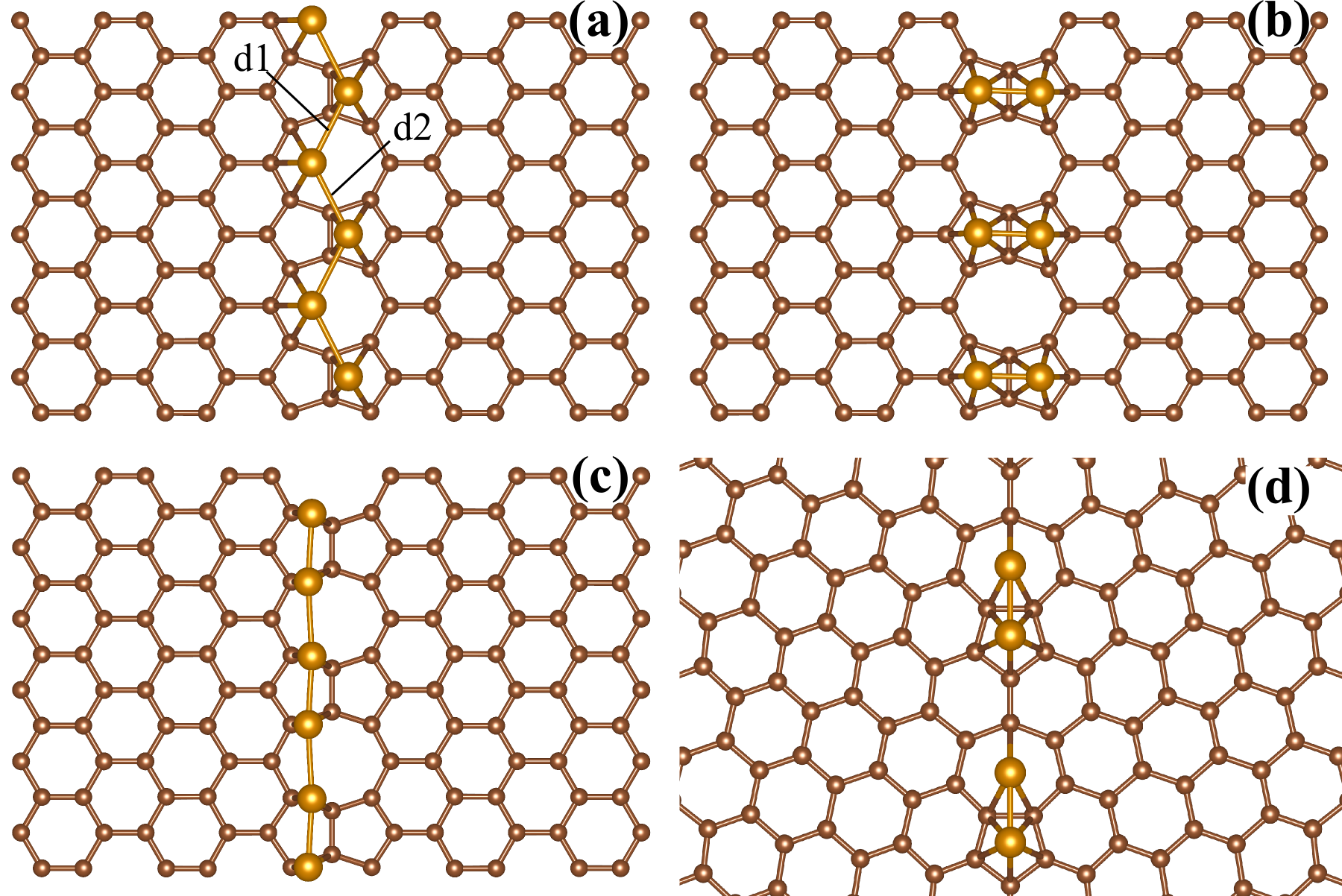}
 \caption{\label{models2d} Structural models of TM[PO]/GB(5-8) for Fe[PO] and 
Mn[PO] (a), TM[PP]/GB(5-8) (b), TM[PO]/GB(5-8) for Co[PO] and Ru[PO], and 
TM[PH]/GB(5-7) (d).}
 \end{figure}

\subsection{Two TMs per GB unit cell}

We next examine the energetic stability of TM/GB by increasing the concentration 
of TMs along the GB sites. For TM/GB(5-8), we have considered two different 
geometries, \textit{viz.}: one TM adsorbed on the pentagonal ring and another  
on the octagonal ring (TM[PO]), and the two TMs  adsorbed on the two pentagonal 
rings  of GB(5-8) (TM[PP]). In Figs.\,\ref{models2d}(a) and \ref{models2d}(b) we 
present the  calculated  equilibrium geometries of TM[PO]/GB(5-8) and 
TM[PP]/GB(5-8), for  TM = Mn and Fe. As shown in Table~III,  we find that the 
energetic stability of TM/GB(5-8) has been strengthened by increasing the 
concentration of the TMs along the GB sites; being  the TM[PO] configuration 
more stable than TM[PP]. That is, the formation of TM-NLs has been favored upon 
the increase of the TM concentration along the GBs. For instance, Mn[PO]/ and 
Fe[PO]/GB(5-8) are more stable by $\sim$1.2\,eV/atom than their counterparts 
Mn[P]/ and Fe[P]/GB(5-8).  At the equilibrium geometry, we find that Mn[PO] and 
Fe[PO] form a zigzag structure [Fig.\,\ref{models2d}(a)].

\begin{table}[h]
\caption{Adsorption energy ($E^a$ in eV/atom), and the net magnetic moment (m in 
$\mu_B$) of TM/GB(5-8) system, for two TMs per GB unit cell; with one TM 
adsorbed on the pentagonal and another in the  octagonal sites (PO), and the two 
TMs adsorbed on the pentagonal sites (PP).}
\begin{ruledtabular}
\begin{tabular}{ccccc} 
\multicolumn{1}{c}{TM/GB(5-8)} & 
\multicolumn{2}{c}{$E^a$} &
\multicolumn{2}{c}{m}\\
\cline{2-3}\cline{4-5}
        &  PO   & PP  &  PO & PP \\
\hline
Mn        &   2.51 & 2.14 &   0.10 & 0.00 \\
Fe        &   2.93 & 2.86 &   3.14 & 3.17 \\
Co        &   3.07 & 2.57 &   1.46 & 2.14 \\
Ru        &   5.32 & 4.02 &   0.00 & 1.29 \\
 \end{tabular}
 \end{ruledtabular}
 \end{table}
 
The energetic preference to the formation of  TM-NLs along the 
GBs was also verified in  TM/GB(5-7), Fig.\,\ref{models2d}(d). Here, we have 
considered one TM on the pentagonal site, and another on the heptagonal site, 
TM[PH]/GB(5-7).  As shown in Table~IV, the TM adsorption energies increase by 
$\sim$1\,eV/atom when compared with their counterparts TM[P]/ and TM[H]/GB(5-7) 
systems. Those results for TM/GB(5-8) and TM/GB(5-7) provide further support to 
the recent experimental findings, related to the formation of linear  structures 
of TMs  on graphene ruled by 
GBs\,\cite{kebaliEPhysJD2009,kimNatComm2014,yuACSNano2014}. 

 \begin{table}
  \caption{Adsorption energy ($E^a$ in eV/atom), and the net magnetic moment (m 
in $\mu_B$) of TM/GB(5-7) system, for two TMs per GB unit cell, with one TM 
adsorbed on the pentagonal and another in the heptagonal sites (PH).}
 \begin{ruledtabular}
 \begin{tabular}{ccc}
 TM/GB(5-7)        & $E^a$  & m   \\
 \hline
 Mn         & 1.82 & 0.17  \\
 Fe         & 2.32 & 3.04  \\
 Co         & 2.45 & 1.37  \\
 Ru         & 3.92 & 1.61  \\
\end{tabular}
\end{ruledtabular}
\end{table}

\begin{figure}
 \includegraphics[width= 8.5cm]{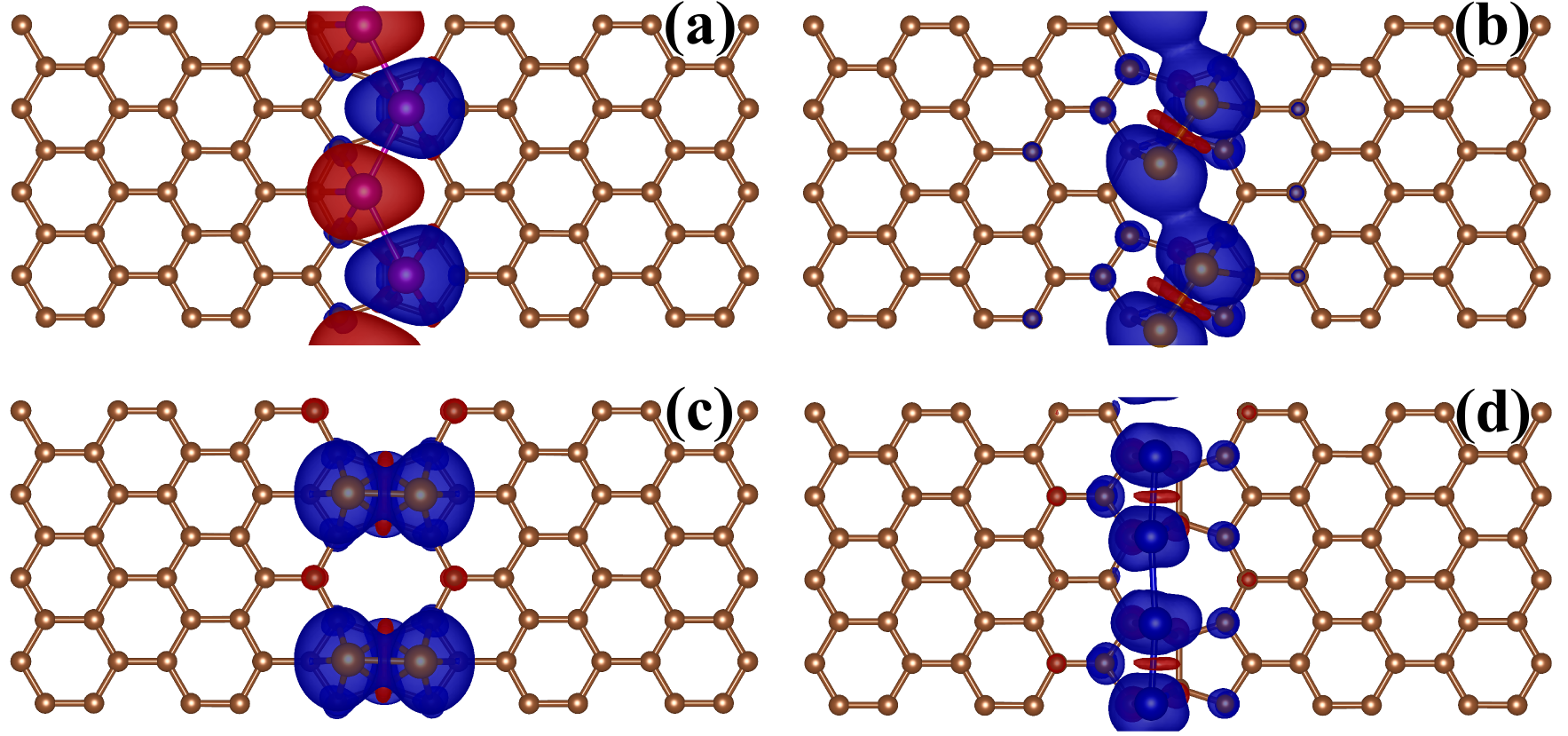}
 \caption{\label{models2e} Spin densities of ferrimagnetic Mn[PO]/GB(5-8) (a),
 FM Fe[PO]/GB(5-8) (b), FM  Fe[PP]/GB(5-8) (c), and ferrimagnetic 
Co[PO]/GB(5-8) (d). Isosurfaces  of $2.2 \cdot 10^{-3}$  $e$/\AA$^3$}
 \end{figure}

The magnetic properties of those TM-NLs   are mediated by  the TM--TM and TM--GB 
interactions. We calculate 
the total energy difference between the AFM and FM states, $\rm\Delta E_{\rm 
AFM-FM} = E_{AFM} - E_{FM}$ for Mn[PO]/GB(5-8). We find that the AFM 
configuration is more stable by $\rm\Delta E_{AFM} - E_{FM} = -0.22$\,eV/atom. 
It is worth noting that the TMs adatoms siting on the P and O sites present 
different hybridizations with the (graphene) host, and thus we have 
different net magnetic moment for TM[P] and TM[O]. In this case, instead of AFM 
coupling,  Mn[PO]/GB(5-8) presents a ferrimagnetic configuration, with a net 
magnetic moment of m = 0.1\,$\mu_{\rm B}$.  In Fig.\,\ref{models2e}(a)  we 
present the spin-density for the  ferrimagnetic Mn[PO]/GB(5-8). At the 
equilibrium geometry, we find Mn--Mn bond length of 2.76\,\AA\ [d1 = d2 = 
2.76\,\AA\ in Fig.\,\ref{models2d}(a)].

In contrast, both Fe[PO]/ and Fe[PP]/GB(5-8)  are FM, with $\rm\Delta E_{\rm 
AFM-FM}$ of 0.14 and 0.21\,eV/atom,  and m = 3.14 and 3.17\,$\mu_{\rm 
B}$/Fe-atom, respectively. Figures\,\ref{models2e}(b) and \ref{models2e}(c)  
present the spin-density  of the FM Fe[PO]/ and Fe[PP]/GB(5-8). At the 
equilibrium  geometry (in the FM state), the Fe adatoms form a dimer-like 
structure, with  a dimer bond length of 2.17\,\AA\ and dimer separation of 
3.36\,\AA,  d1 and d2  in Fig.\,\ref{models2d}(a).  It is worth noting that the  
spin polarization are localized on the Fe adatoms, with almost negligible 
contribution from the host (neighboring) C atoms. Fe[PO]/GB(5-8) and 
Fe[PP]/GB(5-8) are very close in energy, Table~III. The Fe-dimers of 
Fe[PP]/GB(5-8) present a dimer bond length of 2.37 and 2.13\,\AA\ for the FM and 
AFM states, Fig.\,\ref{models2d}(b). By  considering isolated Fe-dimers, we find 
an equilibrium Fe--Fe distance of 1.99  and 2.24\,\AA\ for  the FM  and AFM 
states, where the FM configuration (m = 2.8\,$\mu_{\rm B}$/Fe-atom) is more 
stable by 1.35\,eV than the AFM state, which is in agreement with previous 
theoretical study\,\cite{lebonPRB2008}.   Further comparisons with isolated 
Fe-dimers suggest  that the increase of  the Fe-dimer bond length, when adsorbed 
along the GB sites, contributes to the increase of the net magnetic moment of Fe 
adatoms in Fe[PO]/ and Fe[PP]/GB(5-8), with respect to the isolated Fe-dimers 
(2.8\,$\rightarrow$\,3.2\,$\mu_{\rm B}$)\,\cite{mokrousovPRB2007}. 
  
 \begin{figure}[h]
 \includegraphics[width= 8cm]{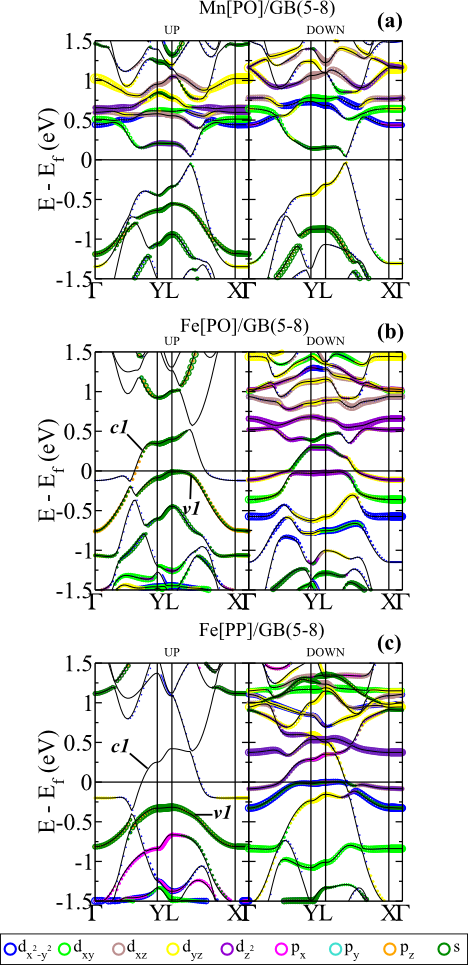}
 \caption{\label{banda2b} Electronic band structure of the ferrimagnetic  
Mn[PO]/GB(5-8) (a), FM Fe[PO]/GB(5-8) (b), and FM Fe[PP]/GB(5-7) (c). The Fermi 
level was set to zero.}
 \end{figure}
 
 \begin{figure}[h]
 \includegraphics[width= 8cm]{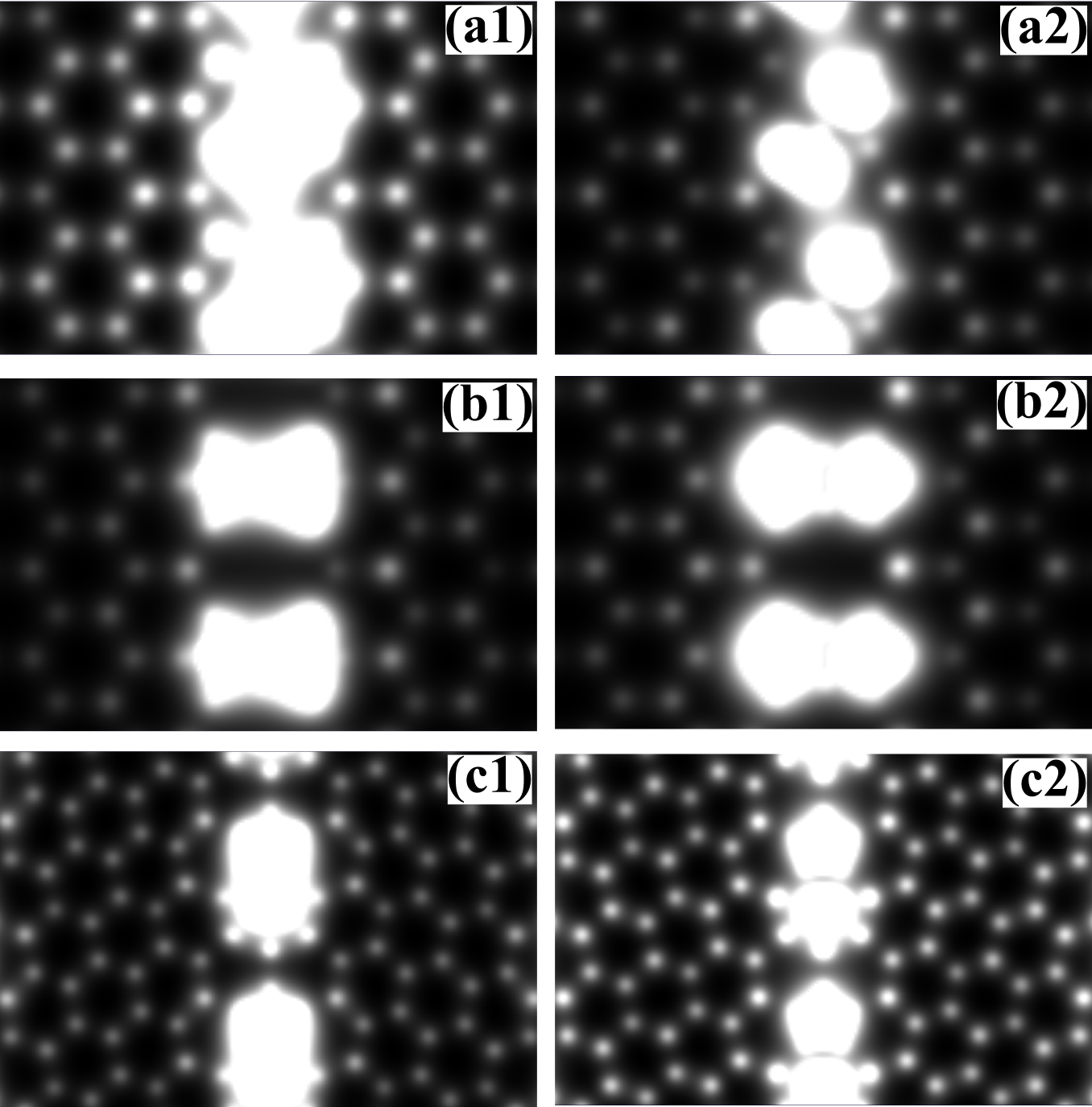}
 \caption{\label{stm} Simulated STM images
of Fe/GB system,  for the occupied  (a1)--(c1), and empty states (a2)--(c2), 
for (a) Fe[PO]/GB(5-8), (b) Fe[PP]/GB(5-8) and (c) Fe[HP]/GB(5-7). We have 
considered an energy interval of 1\,eV with respect to the Fermi level.}
 \end{figure}
 
 \begin{figure}[h]
 \includegraphics[width= 8cm]{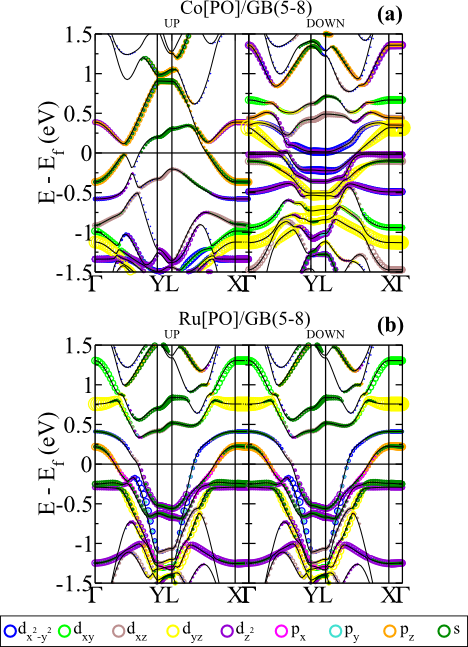}
 \caption{\label{bandaCoRu} Electronic band structure of the ferrimagnetic  
Co[PO]/GB(5-8) (a), Ru[PO]/GB(5-8) (b). The Fermi level was set to zero. }
 \end{figure}
  
The energy bands of spin-up and -down channels of the ferrimagnetic  
Mn[PO]/GB(5-8) system are almost the same, Fig.\,\ref{banda2b}(a). The 
differences are due to the different adsorption sites of Mn adatoms, namely 
Mn[P] and Mn[O]. Here, for both spin channels, we find a Dirac-like structure 
along the L-X direction, as we obtained for the spin-down channels of Mn[P]/ and 
Mn[O]/GB(5-8), Fig.\,\ref{banda-58}(d) and \ref{banda-58}(h). Those Dirac-like 
states are mostly ruled by the graphene-$\pi$ orbitals, with small contributions 
from the Mn-4s and -3d states.  The magnetic moment of each Mn adatom,  m=4.08 
and 4.18\,$\mu_{\rm B}$ for Mn[P] and Mn[O], respectively, are mostly due to the 
unpaired Mn-3d electrons; where  the occupied (unoccupied) Mn-3d orbitals lie at 
about $\rm E_F - 3$\,eV ($\rm E_F  + 1$\,eV).

In Figs.\,\ref{banda2b}(b) and \ref{banda2b}(c) we present the electronic band 
structure of the FM Fe[PO]/ and Fe[PP]/GB(5-8) systems. Near the Fermi level, 
the  electronic band structure of the pristine GB(5-8) has been somewhat 
preserved for the spin-up channel of both systems. In Fe[PO]/GB(5-8), the 
electronic states near the edge of the Brillouin zone  are characterized by 
Fe-4s orbital hybridized with the $\pi$ orbitals of the graphene layer, energy 
band $c1$ in Fig.\,\ref{banda2b}(b). Whereas the electronic contribution of the 
Fe adatoms in $c1$ is negligible for Fe[PP]/GB(5-8), Fig.\,\ref{banda2b}(c); 
indicating that the (spin-up) electronic transport in Fe[PP]/GB(5-8) will be 
mediated by the $\pi$ orbitals of graphene. For both systems, the DP lies below 
the Fermi level, suggesting a net  electronic charge transfer to the graphene 
sheet (n-type doping). In contrast, the spin-down channels are characterized by  
nearly dispersionless  energy bands predominantly composed by Fe-3d orbitals, 
and the Dirac cone feature has been washed out. Thus, similarly to the previous 
Fe[P]/ and Fe[O]/GB(5-8) structures, we may expect a spin-anisotropy on the 
electronic transport along the FM Fe/GB(5-8) system. On the other hand,  such a 
spin-anisotropy has been suppressed in AFM  Fe[PO]/ and Fe[PP]/GB(5-8), where we 
find semiconductor systems (not shown). Indeed, spin-depend electronic current,
perpendicular to the GB sites, has been proposed for other TM geometries on 
Gb(5-8)\,\cite{yuJChemMatC2014}.

In Fig.\,\ref{stm}(a) and \ref{stm}(b) we present the simulated STM images of 
Fe[PO]/ and Fe[PP]/GB(5-8) within an energy interval of $\pm$1\,eV with respect 
to the Fermi level. In Fe[PO]/GB(5-8), the occupied states ($\rm E_F - 1$\,eV) 
are characterized by the formation of a bright line along the  Fe adsorbed GB 
sites [Fig.\,\ref{stm}(a1)], whereas for the empty states  ($\rm E_F + 1$\,eV) 
we find bright spots lying on the Fe adatoms, Fig.\,\ref{stm}(a2). The STM 
pictures for the occupied and empty states [Figs.\,\ref{stm}(b1) and 
\ref{stm}(b2)] are quite similar for Fe[PP]/GB(5-8); characterized by the bright 
Fe dimers perpendicularly to the GB line, which is in agreement with the 
localized feature of the Fe orbitals in Fe[PP]/GB(5-8). The formation of bright 
Fe dimers has been also verified in Fe/GB(5-7), Figs.\,\ref{stm}(c1) and 
\ref{stm}(c2). Somewhat similar STM pictures are expected to the other TM/GB 
systems.  Here, the formation of bright lines along the GB sites, adsorbed by 
TMs, is in accordance with the recent experimental STM 
images\,\cite{kebaliEPhysJD2009,yuACSNano2014,kimNatComm2014}.
 
The equilibrium geometries of Co[PO] and Ru[PO]/GB(5-8) are different than those 
obtained for Mn[PO] and Fe[PO]/GB(5-8). As shown in Fig.\,\ref{models2d}(c), (i) 
the Co (Ru) adatoms form a nearly linear Co-NL (Ru-NL) with d1 = 2.36\,\AA\ and 
d2 = 2.55\,\AA\ (d1 = 2.44\,\AA\ and d2 = 2.48\,\AA), and (ii) the Co and Ru 
adatoms lie on the C--C bridge sites along the GB.  Co[PO]/GB(5-8) present 
a FM state; where the spin-density  is  mainly localized on the Co 
adatoms, Fig.\,\ref{models2e}(d).  Similarly to the other Fe/GB systems, the 
electronic band structure of Co[PO]/GB(5-8) [Fig.\,\ref{bandaCoRu}(a)] suggests 
a spin dependent electronic current. The electronic transport along the Co-NLs  
will be ruled by the spin-up  channels, since the the spin-down electrons will 
face  scattering processes due to the Co-3d localized states near the Fermi 
level. As shown in Fig.\,\ref{bandaCoRu}(a), (i) the energy bands on the spin-up 
channel (near the Fermi level) are characterized by  a strong hybridization 
between the 3p$_{\rm z}$  and 4s orbitals of the Co adatoms with the $\pi$ bands 
of the graphene host. Whereas, (ii) for the spin-down channels, we find a set of 
flat bands,  mostly composed by localized  Co-3d$_{\rm z^2}$ and -3d$_{\rm x^2 - 
y^2}$ states.  It is worth noting that (i) and (ii) were already verified in 
Co/GB systems composed by   a single Co adatoms per grain boundary unit cell, 
Figs.\,\ref{banda-58} and \ref{banda-57}. Here, we find that the formation of 
those dispersionless energy bands near the Fermi level [(ii)] has been 
strengthened, by increasing the concentration of Co adatoms along the GB sites. 
Meanwhile, the Dirac cone structure has been suppressed for both spin-channels 
in     Ru[PO]/GB(5-8), Fig.\,\ref{bandaCoRu}(b).  The electronic band 
structure of Ru[PO]/GB(5-8) can be characterized by the formation of metallic 
bands along the $\Gamma$-Y and L-X direction, composed by Ru-4d and graphene 
$\pi$ orbitals.

 \begin{figure}
 \includegraphics[width= 7cm]{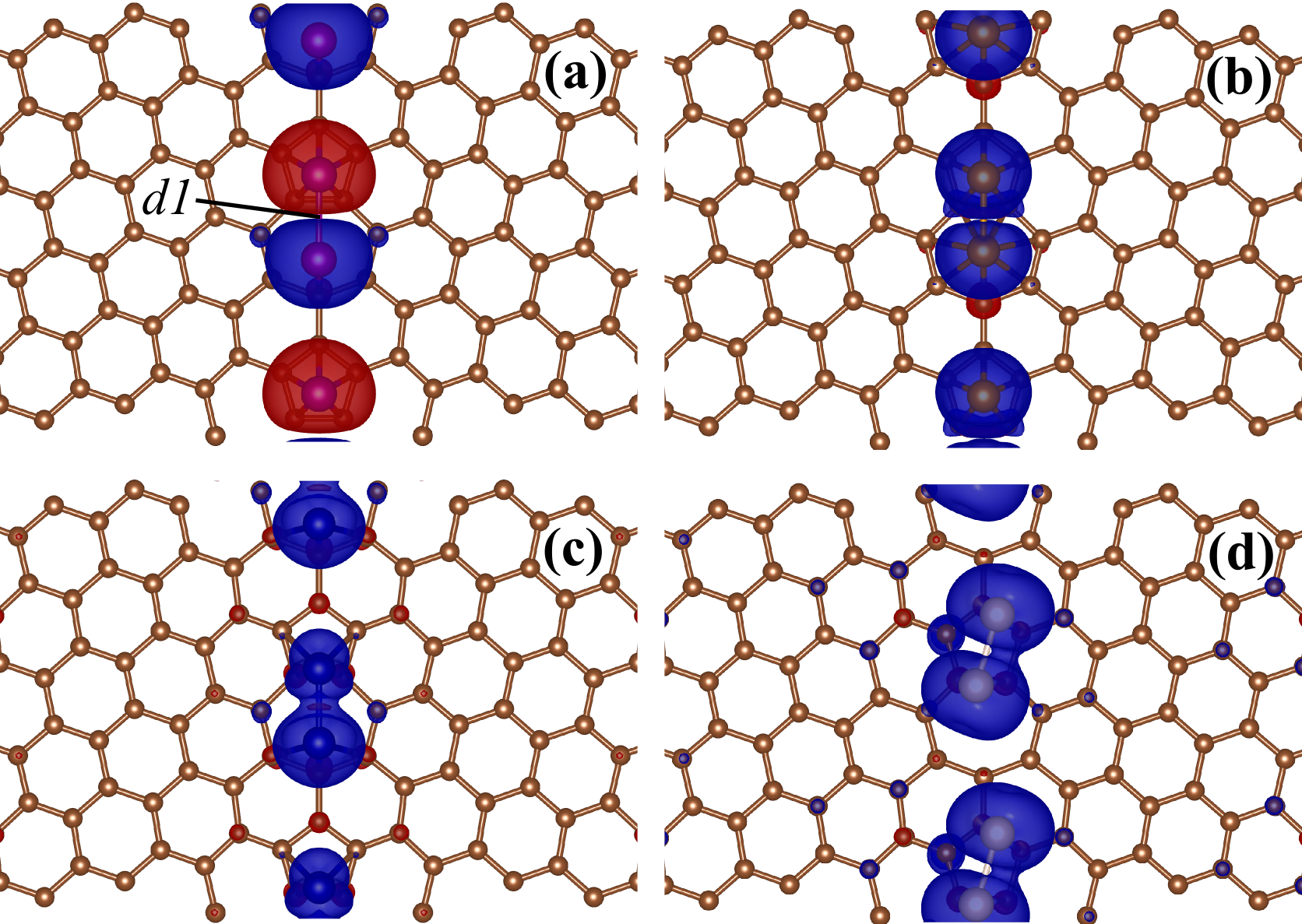}
 \caption{\label{models3} Spin densities of TM[PH]/GB(5-7) for the 
ferrimagnetic Mn[PO]/GB(5-7) (a), FM Fe[PH]/GB(5-7) (b), FM Co[PH]/GB(5-7) (c), 
and FM Ru[PH]/GB(5-7). Isosurfaces of $3 \cdot 10^{-3}$ $e$/\AA$^3$.}
 \end{figure}

 \begin{figure}
 \includegraphics[width= 6.7cm]{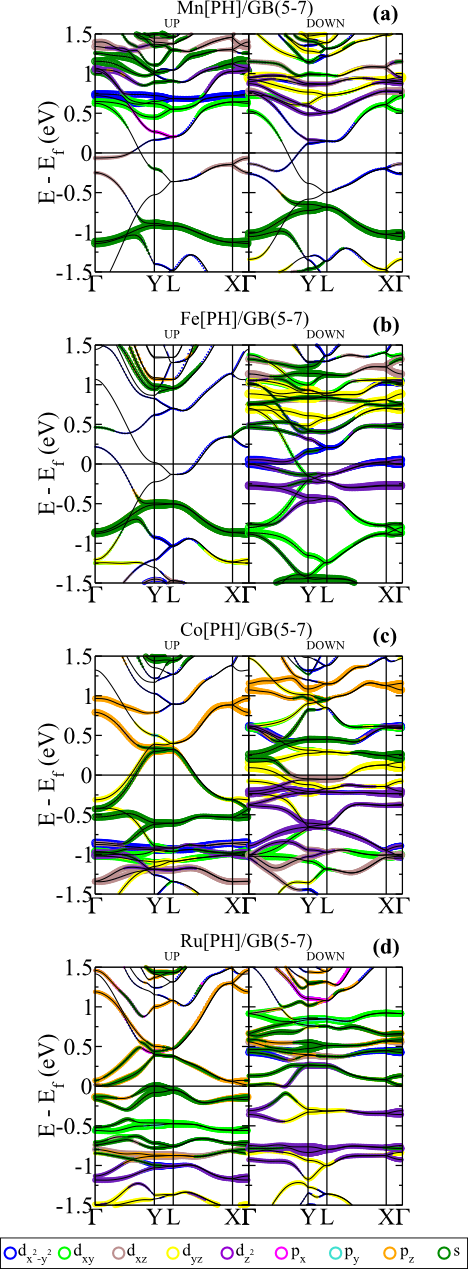}
 \caption{\label{banda-57-2} Electronic band structures  of TM[PH]/GB(5-7) for 
the  ferrimagnetic Mn[PO]/GB(5-7) (a), FM Fe[PH]/GB(5-7) (b), FM Co[PH]/GB(5-7) 
(c), and FM  Ru[PH]/GB(5-7) (d). The Fermi level was set to zero.}
 \end{figure}
 
The Mn[PH]/GB(5-7) system presents a  ferrimagnetic state as the  lowest energy 
configuration. As shown in Fig.\,\ref{models3}(a), the spin densities on the 
Mn[P] and Mn[H] adatoms are slightly different, giving rise to a net magnetic 
moment of 0.16\,$\mu_{\rm B}$. The magnetic moment of each Mn adatom,   m=4.30 
and 4.25\,$\mu_{\rm B}$ for Mn[P] and Mn[O],  are mostly due to the unpaired 
Mn-3d electrons; with  the occupied (unoccupied) Mn-3d orbitals lying at about 
$\rm E_F - 3$\,eV ($\rm E_F  + 1$\,eV). In Fig.\,\ref{banda-57-2}(a) we present 
the electronic band structure of Mn[PO]/GB(5-7). We find, for both spin 
channels, metallic bands for wave vectors along the $\Gamma$-Y direction, 
composed by Mn-3d$_{\rm x^2-y^2}$ and -3d$_{\rm xz}$ states hybridized with the 
host $\pi$ orbitals.

In contrast, the other TM[PH]/GB(5-7) systems (TM = Fe, Co and Ru) are more 
stable at the FM state. In Fig.\,\ref{models3}(b)--\ref{models3}(d) we present 
the (ground state) spin-densities of Fe[PH]/, Co[PH]/,and Ru[PH]/GB(5-7); where 
we find the formation of dimer-like structures along the GB sites. However, the 
electronic structure of those TM[PH]/GB(5-7) systems are quite different. In 
Fe[PO]/GB(5-7) [Fig.\,\ref{banda-57-2}(b)], the  electronic contribution of Fe 
adatoms for the spin-up metallic states is  negligible, whereas for the 
spin-down channel, the Fe-3d orbitals give rise to  dispersionless energy bands 
near the Fermi level. In contrast, in Co[PH]/GB(5-7),  the Co-4s and -3d$_{\rm 
yz}$ orbitals contribute to the formation of (spin-up) metallic bands along the 
$\Gamma$-Y and L-X directions [Fig.\,\ref{banda-57-2}(c)]; indicating a charge 
density overlap,  of Co-4s and -3d$_{\rm yz}$ orbitals,  between the nearest 
neighbor Co-dimers along the GB sites. Whereas,  for the spin-down energy bands, 
we find a set of dispersionless energy bands ruled by the Co-3d orbitals. Such a 
charge density overlap is somewhat reduced for  the Ru-dimers in Ru[PH]/GB(5-7). 
As shown in Fig.\,\ref{banda-57-2}(d), there are no  metallic bands for both 
spin channels of FM Ru[PO]/GB(5-7). The electronic bands near the Fermi level 
present an energy dispersion of $\sim$0.3\,eV, and are mostly composed by Ru-5s 
and -4$d_{\rm z^2}$ orbitals. 

In summary, our results show that the 
formation of TM-NLs ruled by GBs in graphene is a energetically favorable 
process; where the electronic struture of those  TM-NLs is constrained by (i)  
the equilibrium geometry of the TM adatoms (for instance, Fe[PP]/ and 
Fe[PO]/GB(5-8)]), well as (ii) the atomic  geometry of the GB sites like in 
Fe[P]/GB(5-8) and Fe[P]/GB(5-7).

\section{Conclusions}

Based on the {\it ab initio} DFT calculations, we find 
energetically stable TM nanolines in graphene ruled by GBs (TM = Mn, Fe, Co, and 
Ru). The stability of those TM nanolines is strengthened upon TM--TM chemical 
interactions. Our total energy results provide further support to the 
recent experimental findings of TM nanolines in graphene patterned by GBs. At 
the equilibrium geometry, we find ferromagnetic (ferrimagnetic) Fe and 
Co (Mn) nanolines, while the magnetic state of the Ru nanolines depends on the 
(local) adsorption geometry. Through extensive electronic band structure 
calculations, we verify that  the most of TM nanolines are (i) metallic, and  
(ii) spin-polarized with quite different spin-up and spin-down electronic band 
structures. In (ii) we can infer a spin-anisotropy for the electronic current 
along those  TM nanolines.
\acknowledgments

The authors acknowledge financial support from the Brazilian agencies CNPq/INCT 
and  FAPEMIG,  and the computational support from CENAPAD/SP.

\bibliography{../RHMiwa}

\begin{thebibliography}{38}
\bibitem [{MnG()}]{MnGB}%
  \BibitemOpen
  \href@noop {} {}\bibinfo {note} {We have used a energy cutoff of 35~Ry for
  the plane wave basis set, and 280~Ry to describe the total charge density.
  The Brillouin zone was sampled by using a grid of
  1$\times$10$\times$1.}\BibitemShut {Stop}%
\end{thebibliography}

\end{document}